\documentclass[11pt]{article}
\usepackage{graphicx}
\usepackage{epstopdf}
\begin{document}
\title{ New Bright Carbon Stars Found In The DFBS. \\
(Research note, submitted to Astrophysics)}
 \author{
 K. S. Gigoyan\\ 
\small{$^1$V. A. Ambartsumian Byurakan Astrophysical Observatory, Armenia,  email:  kgigoyan@bao.sci.am }
  \and
 C. Rossi, S. Sclavi, S.Gaudenzi\\
\small{$^2$Department of Physics, University La Sapienza, Piazza A.Moro 00185, Roma, Italy}
               }
\maketitle

\section{Introduction}                                 
      
   Carbon Ð rich stars of Population II, such as CH giants, can provide        direct information on the role of low Ð to Ð intermediate Ð mass stars of the Halo in early Galactic evolution. Moreover, accurate knowledge of the CH stellar population is a critical requirement for building up scenarios for early Galactic chemical evolution. The first list of the faint high Ð latitude  C  stars( FHLCs ), found in the Digitized First Byurakan Survey \footnote{http://byurakan.phys.uniroma1.it and  http://www.aras.am/Dfbs/dfbs.html  }
   (DFBS [1 ] ) is given in paper [ 2 ]. In the present work, we report  the  recent discovery  of  two  additional CH Ð type C stars (not previously catalogued), namely DFBS J075331.98+190344.3 and DFBS J111422.94+091442.7,  detected on the DFBS plates with help of the image analysis softwares ( FITSView and  SAOImage DS9 ).  Medium Ð resolution spectra confirm  the C Ð rich  nature for both of them. Using  infrared   color Ð magnitude  relationship, we estimated the distances and   K Ð band absolute magnitudes to the new objects.         
      
      \section{ Optical Spectroscopy And Photometry}
      
\begin{figure*}
   \centering
\includegraphics[width=\textwidth]{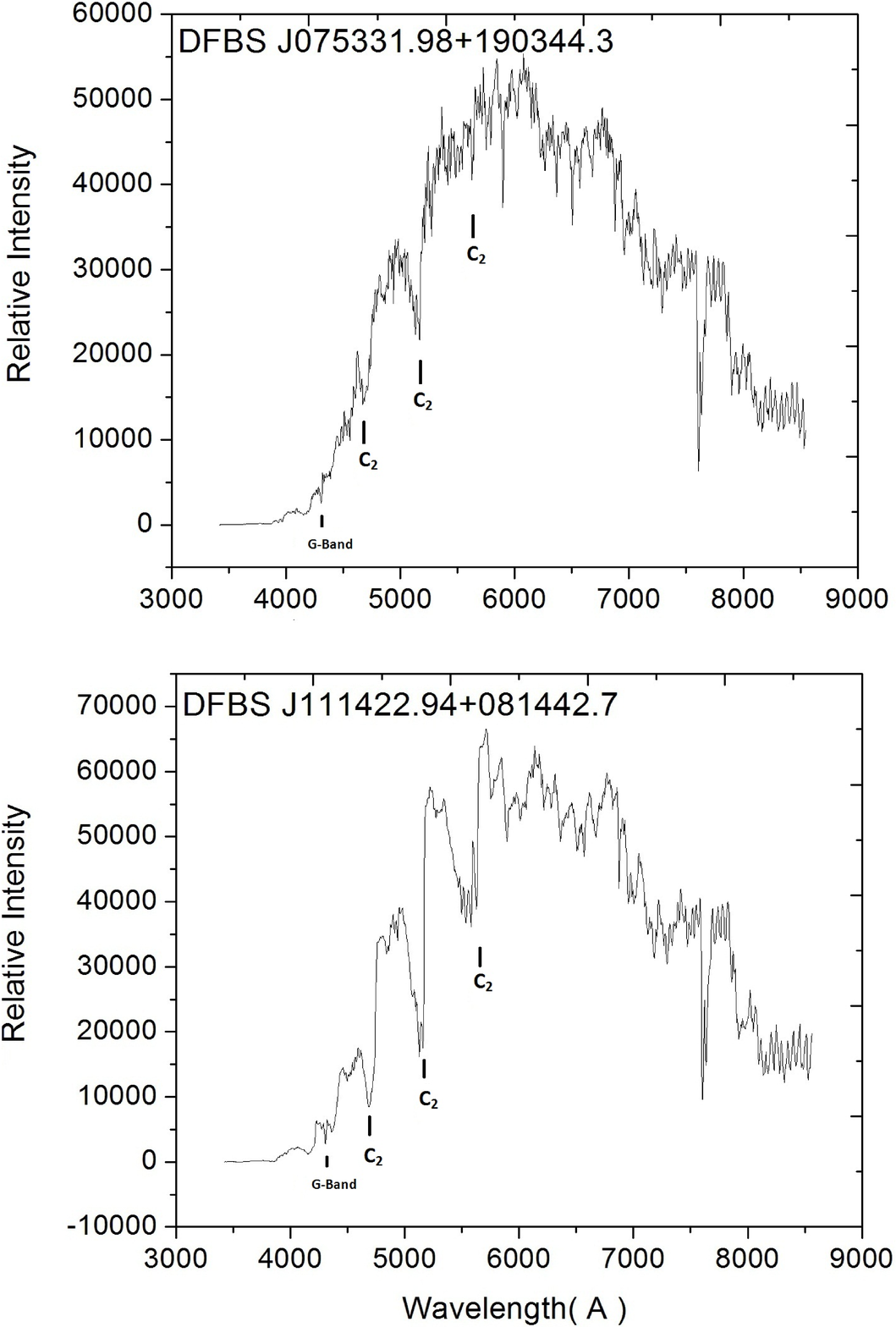}
 \caption{ Medium resolution CCD spectra in the range  3500-8500\AA for the new detected DFBS C stars. The absorption band heads of  the C2 molecule and  G-band of the CH molecule is indicated. The Y axis   is intensity in relative units.  }
 \label{fig1}
 \end{figure*}

       For our stars follow Ð up photometry ( Johnson B,V, R ) and moderate Ð resolution CCD spectra ( spectral  range 3900-8500\AA, dispersion 3.9 \AA /pix )  were obtained on 12/13 March 2012, with the 1.52 m Cassini telescope of the Bologna ( Italy) Astronomical Observatory at Loiano (equipped with the Bologna Faint Object Spectrometer and Camera Ð BFOSC, 1300x1340 pix,EEV P129915 CCD ). All the spectroscopic and photometric data were reduced by means of standard IRAF
 \footnote{ IRAF is distributed by the NOAO which is operated by AURA under contract with NFS }
        procedures. For these  stars Table \ref{tab1} presents: the DFBS identification, which includes the equatorial coordinates; the galactic coordinates $l$ and $b$ ;  the B,V,R  magnitudes ( typical errors are ±0.05 mag );  the spectral class ( see chapter 3 ) determined from the CCD spectra and the value of E(B-V) along the line of sight to the stars, computed using the Galactic reddening maps of Schlegel et al.[ 3 ]. The spectra are shown in Fig. \ref{fig1}, where on Y Ð axis we plotted relative fluxes, corrected for the atmospheric extinction.

 \medskip                    
  \begin{table} []
\caption{The Journal Of  Observations For  The  New  DFBS  C  Stars. }            
\smallskip 
\label{tab1}      
\centering          
\begin{tabular}{c c c c c c  c c   }     
\hline\hline       
 DFBS &   $l$ & $b$ &$B$ &$V$ &$R$ &Sp.type &E($B-V$)\\
       Number &deg &deg& mag& mag& mag & & mag \\
  \hline                  
J075331.98+190344.3&	 202.0349 &+21.9656&	13.36& 11.67& 10.98& CH&  0.044\\
J111422.94+081442.7&	247.8378& +60.2703&	12.96&11.69&	11.14&   CH & 0.031\\
 \hline                  
\end{tabular}
\end{table}
\medskip        
                                                                     
  \section{ Spectral Types And Characteristics}
   The new  data  were  analyzed to clarify the subclass of the  new C stars. The spectra  show strong G-band  of CH molecule at 4300\AA,  which is a main spectroscopic characteristic feature of CH-type stars[ 4, 5 ]. Also, they show the secondary P- branch of the G-band ( with head at 4342\AA ), clearly indicating the belonging of these objects to the class of CH giants [ 6 ]. Prominent features of the C2 molecule at 4737, 5165, 5636\AA,  those in the region 6000 Ð 6200\AA, the $^{13}$CN band near 6360\AA,  and the atomic lines 4554 and 4935\AA  ~of BaII  are very well expressed.    
     Near infrared photometric data were also considered for the new C stars.  Table \ref{tab2} presents the 2MASS magnitudes( available online at http://irsa.ipac.caltect.edu ) and the  $J-H$  and  $H-K $ colors,  transformed to the SAAO  photometric system according to the formulae  by  Koen et al, [ 7 ]  and corrected for the interstellar extinction according to [3].  The uncertainties are 0.040 and 0.046 mag for the colors of  J075331.98+190344.3 and J111422.94+081442.7, respectively. In the $J -H$ vs.  $H-K$ diagram of  Fig.3 by Totten et al. [ 8 ], where the different  carbon classes were established, the colors of the two stars  are typical for CH-type C stars, confirming the spectral classification (see papers [ 8, 9 ] for more details).
                                  
 \medskip                    
  \begin{table} []
\caption{ 2MASS   Photometric Data For The  New DFBS C Stars   }             
\smallskip
\label{tab2}      
\centering          
\begin{tabular}{c c c c c c  c  }     
\hline\hline                                                                       
  DFBS Number &  	2MASS  Number   &  $J$  &$H $ & $K_S$& $J - H$ & $H - K$\\   
 & & mag& mag& mag &mag &mag \\
\hline
J075331.98+190344.3& J07533198+1903441	& 9.224&  8.462&  8.272	& 0.87& 0.15\\
J111422.94+081442.7& J11142294+0814427	 & 9.544& 8.940&  8.796	 &0.69 &  0.11\\
 \hline                  
\end{tabular}
\end{table}
\medskip                                                                             
                                              
\section{ Luminosities And Distances. }
To compute the absolute magnitudes MK  and the distances 
to the new detected objects we used the  empirical color Ð magnitude relationship:

\medskip
 Log( MK + 9.0 ) = 1.14 - 0.65( J-K ) ~~~~~~~~~~~~~~~~~~~~~~~~~~~~~~~~~~~~~~~~~~~~~ ( 1 )     

 \medskip\par\noindent                                 
obtained by Totten et al.[ 8 ] from a selected sample of  C giants in nearby Galactic satellite systems and successfully applied to all their faint high latitude carbon stars. Table \ref{tab3}  presents the absolute K band magnitudes M$_K$,  in the SAAO system, Heliocentric distances ( D ), and the distance to the Galactic plane( Z ).  
                                                             
  \medskip                    
  \begin{table} [h]
\caption{  Absolute  K Ð band  magnitudes and distances  to the  DFBS C Stars   }        
\smallskip     
\label{tab3}      
\centering          
\begin{tabular}{c c c c   }     
\hline\hline                                                                       
  DFBS Number &  M$_K$ (mag)  &  D(kpc) &Z(kpc) \\   
 \hline
J075331.98+190344.3& - 5.80 $\pm$ 0.2 & 6.3 $\pm$ 0.7 & 2.4 $\pm$ 0.7 \\
J111422.94+081442.7&  - 4.75 $\pm$ 0.2 & 5.2 $\pm$ 0.6 & 4.5 $\pm$ 0.6 \\
 \hline                  
\end{tabular}
\end{table}
\medskip

 \section{Summary} Optical spectra in the range 3500-8500\AA and photometric data for two carbon  
stars found in the Digitized First Byurakan Survey database is presented. Both objects are 
 CH Ð type giants, consequently  at distances 6.3 and 5.2 kpc from the Sun.
 \bigskip  
 
{\small {\bf acknowledgements}
 ~~~  The authors thank the staff  of  the Cassini telescope  for technical assistance during the observations. This research has made use of the SIMBAD database operated at CDS, Strasbourg, France. This publication makes use of data products from 2MASS, which is a joint project of the University of Massachusetts     and the Infrared Processing and Analysis Center, California Institute of Technology, funded by the National  Aeronautics and Space Administration and the National Science Foundation. }
                                                  
   {}
   
     \end {document}